\documentclass{aa}
\input psfig.sty

\newcommand{\nat}[2]{Nat #1, #2}
\newcommand{\apj}[2]{ApJ #1, #2}
\newcommand{\apjs}[2]{ApJS #1, #2}

\newcommand{\aeta}[2]{A\&A #1, #2}

\newcommand{\aetas}[2]{A\&AS #1, #2}
\newcommand{\pasp}[2]{PASP #1, #2}
\newcommand{\mn}[2]{MNRAS #1, #2}

\newcommand{\nh}{N$_{\rm H}$}

\newcommand{\Hbet}{H${\beta}$}

\newcommand{\ergs}{erg s$^{-1}$}
\newcommand{\kms}{km s$^{-1}$}

\newcommand{\Teff}{$T_{\rm eff}$\,}
\newcommand{\Tbb}{$T_{\rm bb}$\,}

\newcommand{\degree}{\degr}
\newcommand{\Lbol}{\rm L$_{\rm bol}$}

\newcommand{\fxfo}{F$_{\rm X}$/F$_{\rm opt}$}
\newcommand{\rqvd}{r$_{90}$}

\newcommand{\rxa}{\object{RX\,J0720.4-3125}}
\newcommand{\rxb}{\object{RX\,J1856.5-3754}}
\newcommand{\rx}{\object{RX\,J1605.3+3249}}
\newcommand{\rxs}{\object{1RXS\,J160518.8+324907}}
\begin{document}

   \thesaurus{06         
              (13.25.3;  
               08.14.1;}  

   \title{The isolated neutron star candidate \rx}

   \author{C. Motch 
      \inst{1}
      \thanks{Visiting Astronomer, Canada-France-Hawaii Telescope operated by
	    the National Research Council of Canada, the Centre National de la
	    Recherche Scientifique de France and the University of Hawaii.}
      \and
      F. Haberl
      \inst{2}
      \and
      F.-J. Zickgraf
      \inst{1}
      \and
      G. Hasinger
      \inst{3}
      \and
      A.D. Schwope
      \inst{3}
          }

   \offprints{C. Motch}

   \institute{
              Observatoire Astronomique, UA 1280 CNRS, 11 rue de l'Universit\'e,
              F-67000 Strasbourg, France
              \and 
              Max-Planck-Institut f\"ur extraterrestrische Physik, D-85740,
              Garching bei M\"unchen, Germany 
	      \and
	      Astrophysikalisches Institut Potsdam,
              An der Sternwarte 16, D-14482 Potsdam           
              }
	      
   \date{Accepted for publication in Astronomy \& Astrophysics }

   \maketitle

   \begin{abstract}

   We report on X-ray and optical observations of a ROSAT X-ray source, \rx ,
   selected from the all-sky survey on the basis of its spectral softness and
   lack of bright optical counterpart. The ROSAT PSPC energy distribution is
   well fitted by a blackbody with kT = 92$\pm$6\,eV and \nh\ =
   1.1$\pm$0.4$\times$10$^{20}$\,cm$^{-2}$.  X-ray observations spanning 6.5
   years fail to reveal any flux or spectral  variability on any time scales. 
   The ROSAT HRI error circle only contains a R = 23.25 M star which is
   unlikely to be associated with the X-ray source. We conclude that \rx \ is a
   probable nearby isolated neutron star detected from its thermal emission.
   The present data do not allow to unambiguously determine the X-ray powering
   mechanism, cooling from a young neutron star or heating by accretion from
   the interstellar medium onto an old neutron star. However, the long term
   stability of the X-ray flux favours the young neutron star hypothesis. 

 
   \end{abstract}

%

\section{Introduction}

Several arguments based on present metallicity of the interstellar medium, rate
of supernovae and on properties of the observed radio pulsar population
indicate that of the order of 10$^{8}$ to 10$^{9}$ old isolated neutron star
(INS) should exist in the Galaxy. Ostriker, Rees \& Silk (\cite{ostricker})
were the first to propose that a sizeable fraction of these old and radio-quiet
neutron stars could be heated by accretion from interstellar medium and be
again detectable by their far-UV and soft X-ray emission. Early modelling of
this population by Treves \& Colpi (\cite{treves91}) and Madau \& Blaes
(\cite{mb94}) led to the conclusion that re-heated old neutron stars should
indeed appear in large numbers in soft and UV all-sky surveys with a possible
concentration in the directions of highest interstellar medium densities,
namely the galactic plane in general and more specifically molecular clouds. 

Boosted by these predictions several optical identification campaigns of ROSAT
X-ray and UV sources were initiated and it readily became clear that the number
of possible isolated neutron star candidates was substantially below average
model predictions (e.g., Manning et al. \cite{manning}, Motch et al.
\cite{motch97}, Danner \cite{danner1}). However, four good isolated neutron
star candidates were discovered so far in the ROSAT all-sky survey. These
candidates share as common properties a soft X-ray spectrum with black body
temperatures below 100\,eV, no radio emission detected and very high \fxfo\
ratio in excess of 10$^{4}$. The X-ray brightest of these candidates are \rxb \
which was optically identified with a V=25.6 blue object (Walter et al.
\cite{walter97}) and the pulsating source \rxa \ (Haberl et al.
\cite{haberl97}) which has no counterpart brighter than B=26.1 (Motch \& Haberl
\cite{mh98}, Kulkarni \& van Kerkwijk \cite{KvK98}). Other good cases are
RX\,J0806.4--4123  (Haberl et al. \cite{haberl98}) and 1RXS J130848.6+212708
(Schwope et al. \cite{schwope}).

In the mean time, the possibility that a fraction of these candidates could
rather be young neutron stars has gained considerable credit.  The lack of
detectable radio emission from these sources could be explained by a B, 
P$_{\rm spin}$ position behind the radio pulsar death line or more simply by
beaming effects which are stronger at long spin periods (Wang et al.
\cite{wang98}). Finally, RXTE observations have demonstrated that soft
$\gamma$-ray repeaters are newly born neutron stars with extremely high
magnetic fields (Kouveliotou et al. \cite{kou98}) and belong to the population
of magnetars proposed by Duncan \& Thomson (\cite{duncan92}). As radio emission
can be quenched by the strong magnetic field, these objects may remain
undetected by classical radio means and their birth rate could amount to 10\%
of that of ordinary pulsars (Kouveliotou et al. \cite{kou94}). Since the
primary store of energy in a magnetar is that in the magnetic field, B decay
could constitute a significant source of heat, allowing magnetars to remain
detectable in X-rays over longer times than ordinary pulsars (Heyl \& Kulkarni
\cite{hk98}). 

In fact, several of the INS found so far could be young neutron stars, perhaps
descendant of soft $\gamma$-ray repeaters, rather than old accreting INS as
originally thought. Detailed study of the few known INS and determination of
their X-ray powering  mechanism is therefore of high importance. 

In this paper, we report on X-ray and optical observations of one of the X-ray
brightest isolated neutron star candidates. The source \rx , also known as RBS
1556 (Schwope et al. \cite{schwope}), was extracted from the ROSAT all-sky
survey on the basis of its soft spectrum and lack of bright optical and radio
counterpart.  

\section{Selection of isolated neutron star candidates from the ROSAT all-sky
survey}

Thanks to its soft sensitivity well suited to the detection of sources with
\Tbb = 20-100 eV, as expected from young cooling neutron stars or from old ones
re-heated by accretion, the ROSAT all-sky survey offers a highly valuable
database for detecting these elusive objects. In order to find candidates, we
selected ROSAT all-sky survey sources displaying HR1 and HR2 hardness ratios
compatible with intrinsically soft spectra slightly modified by a reasonable
amount of interstellar absorption. Hardness ratios 1 and 2 are defined as
\begin{displaymath}
{\rm HR1}\ = \frac{(0.5-2.0)-(0.1-0.4)}{(0.1-0.4)+(0.5-2.0)} \ 
\end{displaymath}
\begin{displaymath}
{\rm HR2}\ = \frac{(1.0-2.0)-(0.5-1.0)}{(1.0-2.0)} \
\end{displaymath}
\noindent where (A-B) is the raw background corrected source count rate in the A$-$B
energy range expressed in keV. Based on simulations of black body energy
distributions folded with the ROSAT PSPC response we decided to extract all-sky
survey sources having hardness ratios compatible (i.e. within one standard
error value) with HR1 $\leq$ -0.25 and HR2 $\leq$ -0.5. This parameter space
corresponds to \Tbb \ $\leq$ 100 eV and \nh \ $\leq$ 4$\times$10$^{21}$ to
3$\times$10$^{20}$\,cm$^{-2}$ for \Tbb\ = 40\,eV and  \Tbb\ = 100\,eV
respectively. Our observational strategy was then to identify all sources in
large sky areas down to the faintest X-ray flux level possible  in order to
find new candidates and also efficiently constrain the space density of these
objects. Results from this global study will be presented in a later paper. The
possible INS nature of \rx \ = \rxs \ was discovered while identifying the
northern sample. \rxs \ has a count rate of 0.875$\pm$0.041 cts/s, HR1 =
$-$0.70$\pm$0.03 and HR2 = $-$0.58$\pm$0.10.  To our knowledge, \rx \ is the
brightest isolated neutron star candidate in the northern hemisphere.

\section{ROSAT observations}

ROSAT observed the field of \rx \ in pointed mode on two occasions. The first
observation was carried out during the 1998 PSPC revival period from 1998
February 18 till 22 for a total exposure time of 4413\,s. The second
observation was performed with the HRI from 1998 March 2 to 4 and lasted
19307\,s. The source was detected on both occasions. 

\subsection{X-ray spectral analysis}

\begin{figure}
\psfig{figure=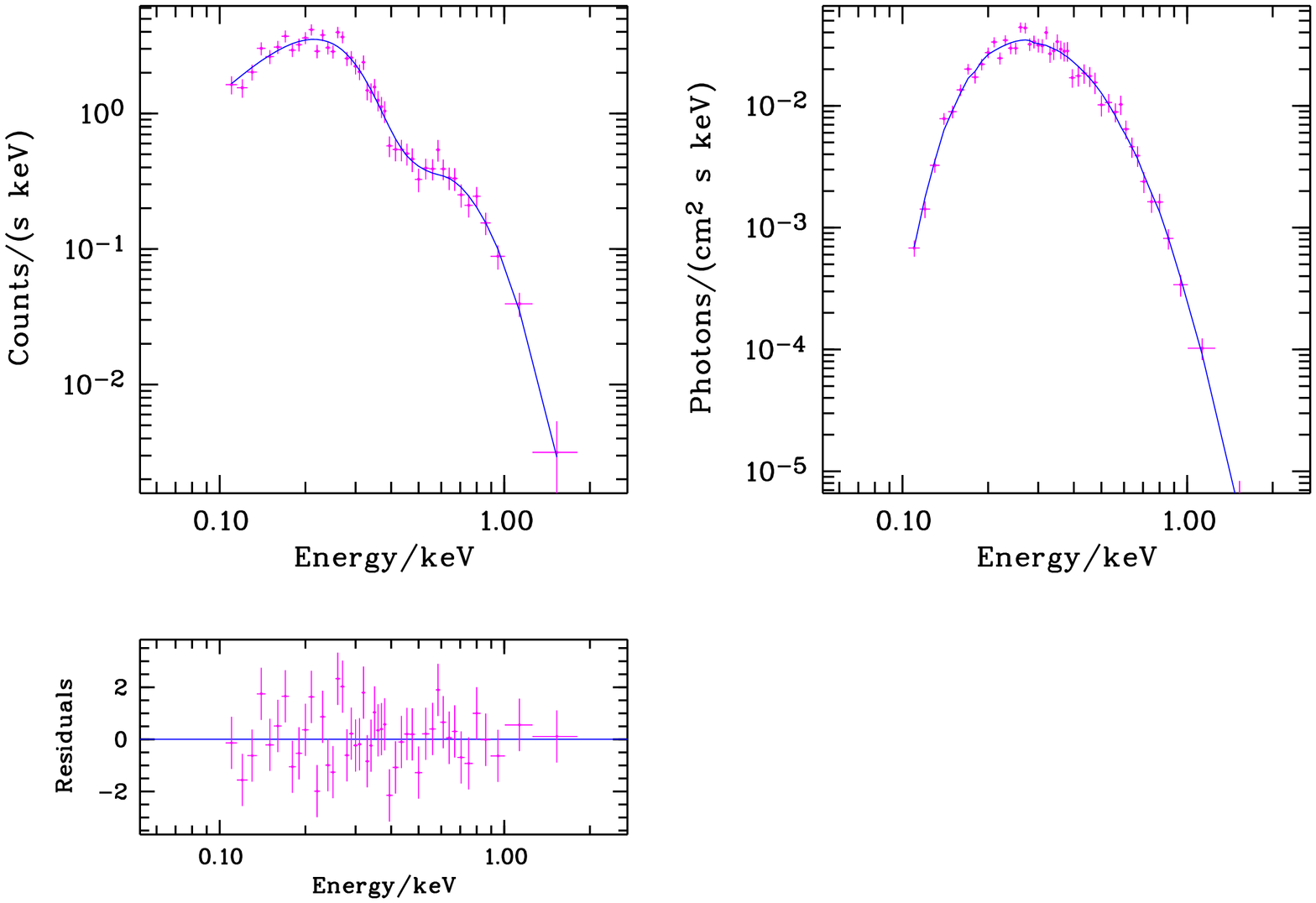,width=8.8cm,bbllx=2.0cm,bburx=20cm,bblly=1.0cm,bbury=14cm,clip=true,angle=-90}
\caption[]{Best blackbody fit to the PSPC count distribution of \rx \ with \Tbb
= 92\,eV and \nh\ = 1.1 10$^{20}$\, cm$^{-2}$}
\label{bbfit}
\end{figure}

\begin{figure}
\psfig{figure=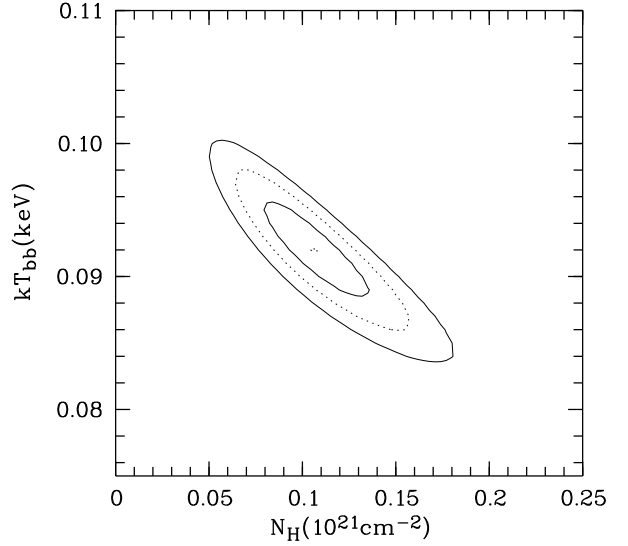,width=8.8cm,bbllx=1.0cm,bburx=18cm,bblly=3.0cm,bbury=18cm,clip=true}
\caption[]{$\chi^{2}$ contour plot showing the allowed parameter space for \nh \ and \Tbb at the
1, 2 and 3 $\sigma$ confidence levels} 
\label{bbgrid}
\end{figure}

The time averaged PSPC spectrum is well represented by a blackbody energy
distribution with best fit parameters of \Tbb\ = 92\,eV and \nh\ = 1.1
10$^{20}$\,cm$^{-2}$ ($\chi ^{2} _{44}$ = 50.9). At the 95\% confidence level,
the allowed range is \Tbb\ = 86 - 98\,eV and \nh\ = 0.6 - 1.5
10$^{20}$\,cm$^{-2}$. We show in Figs. \ref{bbfit} and  \ref{bbgrid} the best
blackbody fit and the corresponding  allowed spectral parameter range. The
observed flux corresponds to a bolometric luminosity of \Lbol \ = 
1.1$\times$10$^{31}$ ($d$/100\,pc)$^{2}$\,\ergs . Assuming isotropic blackbody
emission, the source radius scales as $R$ = 1.1\,km ($d$/100\,pc). Black body
temperature and line of sight absorption compare well with those
observed from other INS candidates such as \rxb \ (\Tbb\ = 57$\pm$1\,eV, \nh\ =
1.4$\pm$0.1 10$^{20}$\,cm$^{-2}$; Walter, Wolk \& Neuh\"auser \cite{walter96})
and \rxa \ (\Tbb\ = 79$\pm$4\,eV, \nh\ = 1.3$\pm$0.3 10$^{20}$\,cm$^{-2}$;
Haberl et al. \cite{haberl97}).

\subsection{Search for X-ray variability}

Source flux appears remarkably constant over years to weeks time scales. We list in
Table \ref{xintensity} the various source intensity measurements which are all
compatible with a strictly constant emission. For the HRI observation we
computed an equivalent PSPC count rate using the blackbody spectral description
derived from the 1998 pointed PSPC observation. The energy distribution 
measured from hardness ratios did not change either between the almost 6.5
years elapsed from survey to pointed PSPC observations.

\begin{table*}
\caption{Flux measurements}
\label{xintensity}
\begin{tabular}{lllll}
Date          & Mode    & Detector  &  PSPC cts/s        &Note\\ \hline
10-11 Aug 1991& survey  & PSPC      &  0.875 $\pm$ 0.041 & 
                HR1 = -0.70$\pm$0.03  HR2 = -0.58$\pm$0.10 \\
18-22 Feb 1998& pointed & PSPC      &  0.900 $\pm$ 0.021 &                 
                HR1 = -0.66$\pm$0.02  HR2 = -0.68$\pm$0.04 \\
02-04 Mar 1998& pointed & HRI       &  0.925 $\pm$ 0.020 & 
                HRI cnt rate = 0.21 $\pm$ 0.01 cts/s \\ 
\hline
\end{tabular}
\end{table*}

We also searched for both aperiodic and periodic variability in the ROSAT PSPC and
HRI time-series. In all cases we failed to detect any significant variability.
Applied to light curves binned in 100\,s intervals the Kolmogorov-Smirnov test
gives a 95\% confidence upper limit of 43\% on variability amplitude for the
PSPC time series and does not provide useful constraints for the HRI data.
Power spectrum analysis also fails to detect any periodic signal with an
estimated upper limit of 50\% and 36\% full amplitude modulations for periods
longer than 1\,s in the PSPC and HRI data respectively. We note that if \rx \
had exhibited pulsations with an amplitude similar to those seen in \rxa \
(24\% , Haberl et al. \cite{haberl97}), we would not have detected them. 

\subsection{Source position}

Considering the extreme optical faintness of the counterpart any attempt to
identify the source obviously requires as good as possible X-ray localization.
Although the positioning of the X-ray source on the HRI or PSPC instrumental
reference grid can be accurate at the arcsec level for relatively bright
sources such as \rx , the uncertainty on the attitude of the satellite
introduces a dominant 8 - 10 arcsec error. However, if enough identified and
well localized sources are present in the field of view, it is possible to
correct for the unknown attitude error and retrieve the intrinsic accuracy
achievable with the given detector.   

\begin{figure}
\psfig{figure=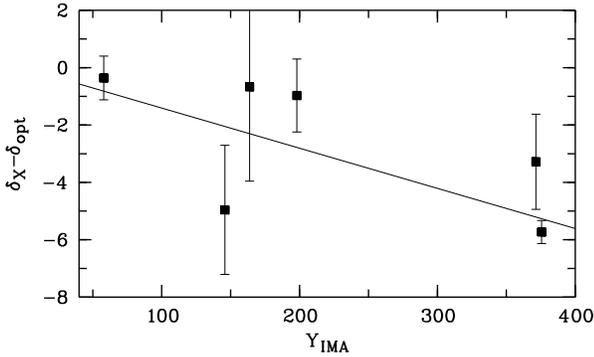,width=8.8cm,bbllx=2.0cm,bburx=18cm,bblly=3.0cm,bbury=13cm,clip=true}
\caption[]{X-ray to optical position offset as a function of HRI pixel number}
\label{hri_cor}
\end{figure}

Above a Maximum Likelihood of 8, a total of 23 and 14 sources are detected in
the PSPC and HRI fields of view respectively. We have cross-correlated the HRI
source list with SIMBAD, FIRST and USNO-A2 catalogues. Among these detections,
4 PSPC and 6 HRI sources have a positive match in the searched catalogues. We
also took into account the 4\arcsec \ shift across the 40\arcmin \ HRI field of
view due to the pixel size being 0.9972$\pm$0.0006 instead of 1 arcsec
(Hasinger et al. \cite{hasinger98}). For one of the HRI and PSPC field source,
RX\,J1605.5+3239, we had two possible identifications, either the nucleus of
the spiral galaxy CASG 1345 or the FIRST radio source located 6.5\arcsec \
away. Identifying the HRI source with the FIRST entry yields attitude
correction vectors incompatible with those derived from other sources in the
HRI field of view. We therefore assumed that the HRI source was identified with
the galactic nucleus. We fitted to the differences between X-ray and optical
positions expressed in arcsec, relations of the form   \mbox{$\alpha _{\rm X}$
$-$$\alpha _{\rm opt}$} = 5\arcsec \, (X$_{\rm IMA}$ $-$ X$_{\rm Center}$) (1
$-$ 0.9972) + Cte and  \mbox{$\delta _{\rm X}$ $-$$\delta _{\rm opt}$} =
5\arcsec \, (Y$_{\rm IMA}$ $-$ Y$_{\rm Center}$) (1 $-$ 0.9972) + Cte where
X$_{\rm IMA}$ and Y$_{\rm IMA}$ are the position of the sources on the grid of
5\arcsec \ size HRI pixels. The fit took into account the error on the
positioning on the instrumental grid. We show in Fig. \ref{hri_cor} the best
fit obtained for the declination axis.

Applying these relations to \rx \ moves the uncorrected HRI position by
0.4\arcsec \ to the East  and 3.6\arcsec \ to the North to $\alpha$ = 16h 05m
18.66s and $\delta$ = +32\degree \ 49\arcmin \ 19.7\arcsec \ (2000.0 eq.) with
a 1 $\sigma$ error of 0.64\arcsec .  Using the 4 identified PSPC sources yields
a compatible position at $\alpha$ = 16h 05m 18.75s and $\delta$ = +32\degree \
49\arcmin \ 17.8\arcsec \ (2000.0 eq.) with a 1 $\sigma$ error of 3.4\arcsec.
The HRI and PSPC pointed positions are well within the 90\% confidence ROSAT
survey error circle ($\alpha$ = 16h 05m 18.8s and $\delta$ = +32\degree \
49\arcmin \ 07.5\arcsec \ (2000.0 eq.) with a 1 $\sigma$ error of 7.0\arcsec).

\section{Optical data}

\subsection{Observations}

First optical observations took place from 1998 April 14 to 17 using the Canada
France Hawaii telescope. Images and spectra were obtained with the OSIS V
instrument which provides image stabilisation and active guiding. With the
2048\,$\times$\,2048 STIS2 CCD the pixel size is 0.156\arcsec \ on the sky. We
acquired  several 15\,min long images through the B, R and I filters. Total
exposure time amounts to 60\,min in B, 75\,min in R and 15\,min in I. FWHM
seeing was quite constant throughout the run with a mean value of 1.0\arcsec .
Images were corrected for bias and flat-fielded using standard MIDAS
procedures. All nights were of photometric quality. Observation of standard
stars in M92 (Christian et al. \cite{christian85}) allowed to calibrate the
images with respect to the Kron-Cousins BVRI system. 

We also obtained low resolution spectroscopy of the 4 brightest optical objects
named A,B,C and D and located inside or close to the ROSAT all-sky survey error
circle. Wavelength calibration was derived from the observation of Hg/Ar arc
spectra and the instrumental response was computed using the flux standard
star Feige 34. The Multiple Object Spectroscopy mode of the OSIS instrument allowed
the simultaneous acquisition of spectra from the 4 objects through 1\arcsec \
slits and using the V150 grism. This configuration yields a FWHM resolution of
$\sim$ 2\,nm and a useful spectral range of 365 to 990\,nm. We acquired 6
individual frames with exposure times ranging from 30 to 60\,min. The total
spectral exposure time is 225\,min. Spectra were flat-fielded, wavelength
calibrated, extracted and corrected for atmospheric absorption and instrumental
response using standard MIDAS procedures. 

Additional images were obtained with the Keck LRIS (Low-Resolution Imaging
Spectrograph) on 1999 February 23. These observations took place under good
sky conditions with a seeing around 0.9 arcsec. Two B and two R images of
the field were taken with a total exposure time of 15\,min in each filter. 
Using observations of one Landolt standard star field, a photometric
calibration with estimated uncertainty of  0.1-0.2 mag was achieved. 

\subsection{Imaging data}

\begin{figure}
\psfig{figure=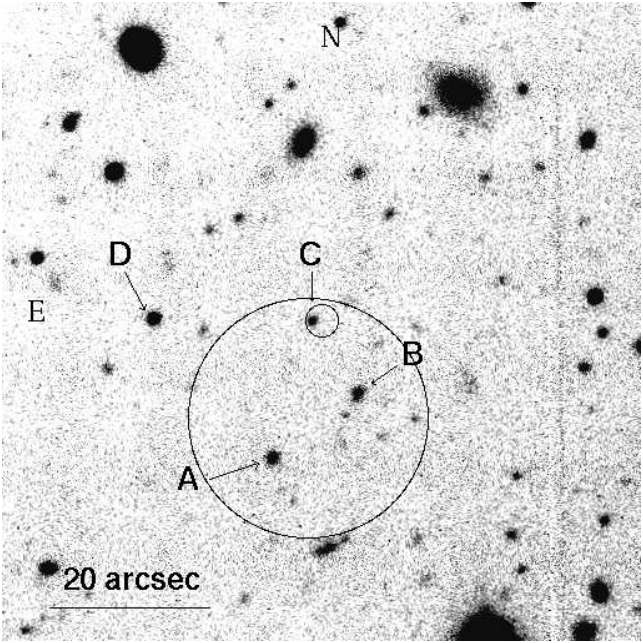,width=8.8cm,bbllx=2.0cm,bburx=18.5cm,bblly=10.0cm,bbury=27cm,clip=true}

\caption[]{The summed CFHT R band image showing the position of the ROSAT HRI
attitude corrected error circle (\rqvd \  = 2\arcsec ), the  ROSAT PSPC survey
error circle (\rqvd \ = 15\arcsec ) and the 4 objects for which we obtained low
resolution spectroscopy.}

\label{fc}
\end{figure}

\begin{figure}
\psfig{figure=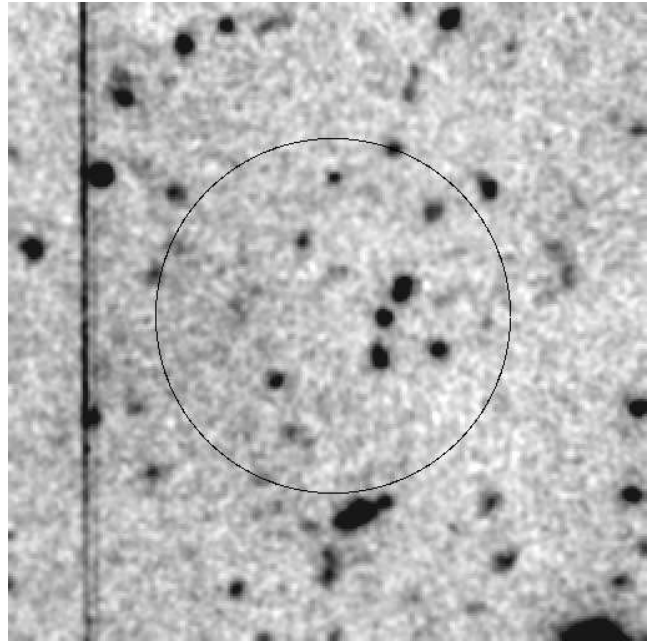,width=8.8cm,bbllx=2.0cm,bburx=18.5cm,bblly=10.0cm,bbury=27cm,clip=true}

\caption[]{The summed Keck B band image with the  ROSAT PSPC survey
error circle (\rqvd \ = 15\arcsec ) overplotted. North is at top and East at left.}

\label{fcb}
\end{figure}

We show on Fig. \ref{fc} the summed CFHT R band image with the 90\% confidence
level ROSAT HRI and ROSAT survey error circles over-plotted. The Keck B image
is displayed in Fig. \ref{fcb}. We calibrated astrometrically our CCD image
using 5 USNO-A2 stellar like objects. The attitude corrected  HRI 90\%
confidence error radius of 2\arcsec \ shown here includes the additional
astrometric error of 0.7\arcsec \ arising from the CCD calibration. The only
detectable object in the ROSAT HRI error circle is C.

R band image profile measurements reveal that objects A and B are resolved. A
seems extended in all directions whereas B has a stellar-like core with diffuse
emission towards the SE direction. On the other hand, objects C and D appear
unresolved. 

BRI photometry of objects A,B,C and D as derived from CFHT observations is
listed in Table \ref{abcd}. Only object D is bright enough to be detected in
the CFHT B band image. We estimate limiting magnitudes of B $\sim$ 24.6 and R
$\sim$ 25.0 on the summed images.

Keck photometry of object C gives R = 23.3 and B-R = 2.6 consistent with
CFHT observations. The limiting magnitudes of the Keck images are estimated to
be B $\sim$ 27 and R $\sim$ 26.

\begin{table}
\caption{BRI photometry of objects A,B,C and D derived from CFHT observations}
\label{abcd}
\begin{tabular}{ccccc}
Star        & B                & R              &  I             & R - I   \\ \hline
A           &  -               & 22.46$\pm$0.04 & 21.51$\pm$0.07 & 0.85$\pm$0.08 \\
B           &  -               & 22.53$\pm$0.04 & 21.03$\pm$0.05 & 1.32$\pm$0.06 \\
C           &  -               & 23.25$\pm$0.05 & 22.03$\pm$0.07 & 1.07$\pm$0.08 \\  
D           &  23.83 $\pm$0.20 & 21.90$\pm$0.03 & 21.34$\pm$0.05 & 0.52$\pm$0.06 \\        
\hline
\end{tabular}
\end{table}

\subsection{Spectroscopic data}

In general, the signal to noise ratio of the average spectra is not good enough
to unambiguously measure a redshift or detect the presence of weak emission
lines. Telluric absorption lines (e.g. the $\lambda$6800-6900\AA \ complex) are
not pronounced.

\begin{figure}
\psfig{figure=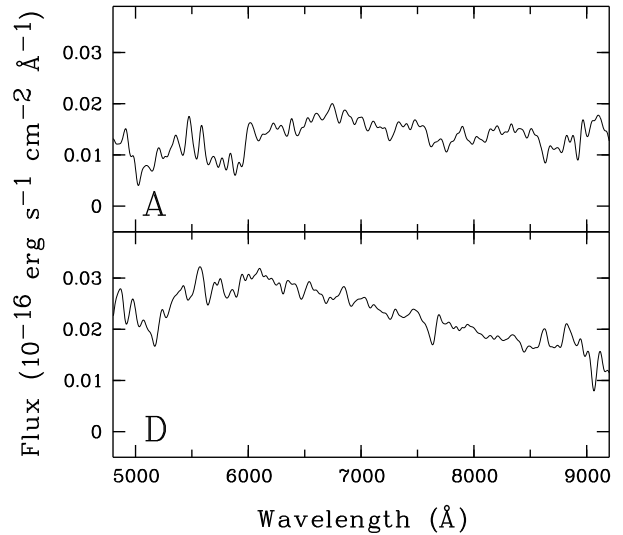,width=8.8cm,bbllx=2.0cm,bburx=18cm,bblly=2.0cm,bbury=17cm,clip=true}
\caption[]{Flux calibrated spectra of objects A and D. 
Spectra have been smoothed by a Gaussian filter with $\sigma$ = 18\AA}
\label{ad}
\end{figure}

Identifying the flux drop bluewards of 6000\AA \ in object A (see Fig.
\ref{ad}) with the Ca break leads to a redshift of $\sim$ 0.5. At this
redshift, other spectral features such as G band, \Hbet\ and Mg band may be seen
in the spectrum. 

\begin{figure}
\psfig{figure=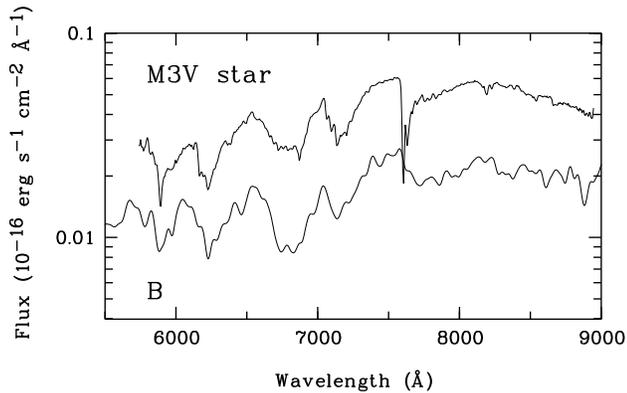,width=8.8cm,bbllx=2.0cm,bburx=18cm,bblly=2.0cm,bbury=13cm,clip=true}

\caption[]{Flux calibrated spectrum of object "B" (bottom) plotted together with that of
a template M3V star (top) . The spectrum of B is  smoothed using a Gaussian filter
with $\sigma$ = 18\AA }

\label{bm}
\end{figure}

The spectrum of object B leaves hardly any doubt that the stellar-like core is
in fact a dwarf M3-4 star, the extended emission being then a likely
background field galaxy. We show on Fig. \ref{bm} the observed spectrum
together with that of a comparison M3V star extracted from the atlas of
Torres-Dodgen \& Weaver (\cite{torres93}). The NaI line, TiO and CaH bands are
clearly detected. Neglecting the contribution of the background galaxy, the
R-I colour index is also consistent with that of a M3-4V star.  

\begin{figure}
\psfig{figure=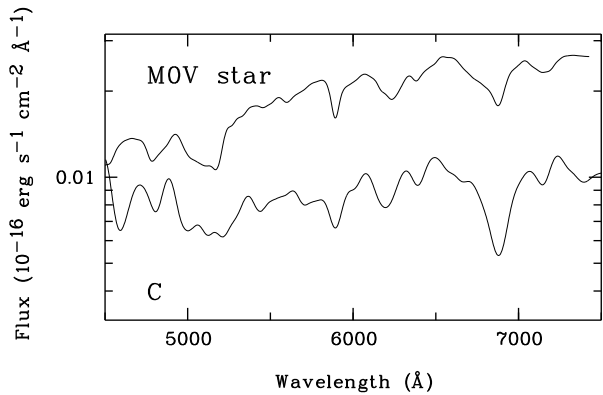,width=8.8cm,bbllx=2.0cm,bburx=18cm,bblly=2.0cm,bbury=13cm,clip=true}
\caption[]{Flux calibrated spectrum of object "C" (bottom) plotted together 
with that of a M0V (top). Both spectra were smoothed using similar Gaussian filters}
\label{cm}
\end{figure}

As object C is the only one detected in the small attitude corrected HRI error
circle, special attention was given to its analysis. Being about 0.7 mag
fainter than star B the spectral features of C are obviously less recognizable.
There is however good evidence that C is also a M star, of slightly earlier
spectral type than star B. For comparison, we show on Fig. \ref{cm} the flux
calibrated spectrum of C together with that of a template M0V star extracted
from the spectral atlas of Jacobi Hunter and Christian (\cite{jhc}). The NaD
line and the broad TiO molecular bands visible in the smoothed M0V spectrum can
be also seen in object C. The similarity of the energy distributions is also
striking and the R-I and B-R colour indexes of C are also consistent with a
$\sim$ M2V star. All these evidences support the conclusion that C is a rather
early M type star. 

Finally, based on Ca break, \Hbet \ and NaD line positions, the spectrum of
object D (Fig. \ref{ad}) suggests an identification with a galaxy at $z$ $\sim$
0.3.

\section{Discussion} 

\subsection{The nature of \rx }

The position of object C in the attitude corrected HRI error circle is
somewhat puzzling but may not be fully significant. At $l$ = 53\degree, $b$ =
48\degree, the surface density of stars brighter than R = 23.5 is 
12,300\,deg$^{-2}$ (Robin \& Cr\'ez\'e \cite{robin86}). There is therefore
more than 1\% probability that a star falls by chance in the 2\arcsec \
radius error circle.  

The photometric distance to the M2V star is $\sim$ 7\,kpc. If the X-ray source
were physically associated with the star, its radius would be $R$ $\sim$
80\,km, i.e. too large for a neutron star and too small for a white dwarf. The
emitting area could be compatible with a polar cap heated by accretion as seen
in AM Her systems in which case, our optical observations could have been
obtained during a low state. This explanation cannot be strictly ruled out, as
the white dwarf could be cool enough in the low state (\Teff $\leq$ 20,000\,K)
to remain undetected at B = 25.85 which is the B magnitude of object C. We also
do not have enough spectral resolution to detect the emission lines which could
reveal heating of the late M star by the white dwarf. However, the absence of
X-ray and optical variability and the unusually hot temperature of \rx \
compared to those of black body components in polars make this possibility
rather unlikely.

A hot white dwarf would have to be located at unrealisticly  large distances
($d$ $\sim$ 270\,kpc), although this distance may be overestimated by the black
body fit. The black body temperature is also hot compared to that observed from
the hottest known PG 1159 stars (e.g. Werner et al. \cite{werner}). Finally, if
the high temperature of the white dwarf were due to nuclear burning at its
surface, we would detect the heated accretion disc and mass donor star or the
surrounding nebula such as in SMC N67.

In general, all classes of soft emitters else than neutron stars are difficult to
reconcile with the observational picture. Taking the expected V magnitude of the M2V
star (V = 24.24) as an upper limit to the optical emission from the ROSAT source
implies \fxfo \ $\geq$ 10$^{4}$. Comparing with the \fxfo \ distribution of bright
ROSAT all-sky survey sources identified in SIMBAD (e.g. Motch et al. \cite{motch98})
shows that all galactic or extragalactic classes of sources other than isolated
neutron stars are probably ruled out (see also Fig. 3 in Schwope et al.
\cite{schwope}). In particular, the source is optically too faint to be identified
with even extreme cases of cataclysmic variables or AGN. 

This conclusion is independent of the HRI attitude correction since none of the
other optical candidates studied in the large ROSAT survey error circle
is a likely counterpart of the X-ray source. In particular, the extragalactic
objects lack the broad emission lines usually seen in soft AGN (e.g. Greiner et
al. \cite{greiner}).

Assuming a neutron star radius of 10\,km implies a source distance of 900\,pc
for full surface emission. However, this distance estimate is very sensitive to
the actual energy distribution in the ROSAT band. For instance, Rajagopal and
Romani (\cite{raro}) have shown that blackbody fits to neutron star model
atmospheres folded through the ROSAT PSPC tend to overestimate the effective
temperature by a factor of up to 3 depending on chemical composition. Applying
the maximum correction could bring the source to much closer distances ($d$
$\sim$ 100\,pc).

The total galactic \nh \ in the direction of \rx \ is 2.47 10$^{20}$\,cm$^{-2}$
(Dickey, J. \& Lockman, \cite{dickey}) about twice that derived from the
blackbody fits to PSPC data (see Fig. \ref{bbgrid}). However, this difference
may not be significant since similar to the effective temperature, the
estimated photelectric absorption sensitively depends upon the assumed soft
X-ray energy distribution.

We conclude that \rx \ exhibits all the features, soft thermal-like spectrum,
high \fxfo , expected from an isolated neutron star.  

\subsection{Accretion from the interstellar medium}

As for \rxa , the average particle density available for accretion in the
vicinity of the neutron star is probably small. IRAS maps do not reveal any
particular density enhancement in the direction of \rx . Using the mean
particle density versus scale height law of Dickey \& Lockman (\cite{dickey})
yields average densities of only 0.3\,cm$^{-3}$ at 100\,pc and 0.06\,cm$^{-3}$
at 300\,pc. In order to explain the X-ray luminosity of the source by Bondi
Hoyle accretion, very low values of the total relative plus sound speed
velocities must be achieved, 12\,\kms \ and 3.5\,\kms \ for $d$ = 100 and
300\,pc respectively. In particular, large distances of the order of 900\,pc as
suggested by PSPC spectral fitting seem incompatible with the accretion model.
However, this major difficulty could be bypassed assuming the presence of a
local overdensity or if the blackbody modelling strongly overestimates the
total accretion luminosity.  

So far, three isolated neutron star candidates have been detected in the
galactic plane ($|b| \leq 20$\degree \ (Haberl et al. \cite{haberl98}) and only
two \rx \ and 1RXS\, J130848.6+212708 (Schwope et al. \cite{schwope}) above the
plane. \rx \ is the X-ray brightest high galactic latitude candidate. The steep
decrease of particle density with distance for \rx \ could allow to
sensitively  test the mechanism leading to X-ray emission in this newly
discovered class of objects. 

\subsection{Cooling neutron star}

The strongest argument in favour of a young cooling neutron star is probably
the stability of X-ray emission over long time scales. This property of \rx \
is also shared by \rxa \ and \rxb . Neutron stars accreting in binaries always
exhibit some kind of variability on a large range of time scales from seconds
to months. Although the accretion conditions prevailing in binaries are
usually far from those assumed for accretion from interstellar medium, it is
still puzzling that none of the members of this new class so far studied in
details shows convincing evidences for variability.

For normal pulsars, the blackbody temperature of 1.1 10$^{6}$\,K implies
cooling ages in the range of 2$\times$ 10$^{4}$ to 10$^{5}$\,yr depending on
the presence of accreted material at the surface of the neutron star (Chabrier
et al. \cite{chabrier}). Considering the probable overestimation of \Teff \ by
the blackbody fit, ages of up to 10$^{6}$\,yr are still possible. 

\rx \ and its cousins \rxa \ and \rxb \ have in common two properties which are
never encountered together in other classes of neutron stars, i) absence of
strong radio emission and ii) absence of luminous hard X-ray tail above the
thermal spectrum.  These features could be used to define the new class of
objects. 

Simultaneous absence of radio and hard X-ray emission is understandable in the
framework of old accreting neutron stars. The low magnetic field or the long
spin period necessary for accretion to take place are likely to put the pulsar
beyond the death line, in the graveyard, and the emitted X-ray spectrum is
expected to resemble that of a soft black body for a large range of parameters
(Zampieri et al. \cite{zampieri}). We envisage below the implications of such
properties for young cooling INS. 

\subsubsection{Radio emission}

The most simple explanation for the absence of radio emission is that the radio
beam does not cross the earth. The beaming fraction which is the proportion of
the sky swept by the radio beam decreases with increasing spin period (Biggs
\cite{biggs}) and is of the order of 0.2 for the overall pulsar population
(e.g. Lyne et al. \cite{lyne98}). Therefore, radio detection means may miss a
large  fraction of the pulsar population among which a part may be hot enough
to be detected in soft X-rays. 

As noted by Kulkarni \& van Kerkwijk (\cite{KvK98}), this explanation cannot
hold for the pulsating source \rxa \ because the time needed to brake the
neutron star to the long spin  period of 8.39\,s (assuming a magnetic field of
10$^{12}$\,G) is much larger than the cooling time. For the same reason, the
absence of radio emission from \rx \ is unlikely to be due to a position beyond
the death line as this would also imply rather long spin periods incompatible
with the hot emission.

Another possibility is that \rx \ is a magnetar with a dipolar surface field B
larger than about 4$\times$10$^{13}$\,G in which case radio emission may be
quenched (Heyl \& Kulkarni \cite{hk98}). 

\subsubsection{X-ray spectrum}

The fact that the X-ray spectrum is to the accuracy of the measurements
thermal-like suggests a much reduced magnetospheric activity compared to other
known neutron stars. 

Among the 27 pulsars detected in X-rays and listed in Becker \& Tr\"umper
(\cite{bt97}), only three middle aged pulsars (Geminga, PSR B0656+14 and PSR
B1055-52) have in addition to a power law a recognizable black body component
in their soft X-ray energy distribution. An interesting case is PSR B0656+14, a
radio emitting pulsar about $\sim$ 10$^{5}$\,yr old with a magnetic field of
4.7$\times$10$^{12}$\,G and located at a distance of 760\,pc. PSR B0656+14
exhibits a ROSAT PSPC spectrum (\Tbb\ = 80-90 eV; Possenti et al.
\cite{possenti}) strikingly similar to that of \rx . The additional faint hard
component needed to fit the spectrum of PSR B0656+14 would not have been
detected in \rx \ because of the lower statistics. 

Neutron stars born with magnetic field B $\geq$ 10$^{14}$\, G, the magnetars,
are thought to be powerful soft X-ray emitters because magnetic field decay
provides an additional source of heat (Thompson \& Duncan \cite{td96}, Heyl \&
Kulkarni \cite{hk98}). The young magnetars associated with soft $\gamma$-ray
repeaters such as SGR 1806-20 or SGR 1900+14 exhibit powerlaw-like quiescent
X-ray spectra without evidences for black body components. These non-thermal
energy distributions could be the signature of a compact synchrotron nebula
(Marsden et al. \cite{marsden}). It has been proposed that the class of
anomalous braking X-ray pulsars could be related to magnetars (e.g.  Thompson
\& Duncan \cite{td96}) and could constitute a later, less active soft
$\gamma$-ray repeater phase. In general, anomalous X-ray pulsars have again
powerlaw-like spectra with in two cases possible blackbody components (see
Thompson \& Duncan \cite{td96} and references therein). Pulsating sources like
\rxa \ could represent an even later stage of magnetar evolution with remaining
very high magnetic field (B $\sim$ 10$^{14}$\, G, Heyl \& Hernquist \cite{hh})
and perhaps still the possibility to emit powerful $\gamma$-ray bursts on
occasion. Because of the additional energy source a magnetar could reach the
temperature of 1.1 10$^{6}$\,K after 10$^{6}$\,yr (Heyl \& Hernquist
\cite{hh}). It is however unclear whether the absence of a strong non-thermal
component in the X-ray spectrum is compatible with the remaining high magnetic
field. 

\subsection{Expected brightness of the optical counterpart}

Taking into account the unfortunate possibility of a chance alignement between
the neutron star and object C implies a B magnitude fainter than $\sim$ 26 for
\rx . Because of the relatively high temperature, the extrapolation of the
black body seen in soft X-rays to the optical regime would imply extremely
faint optical magnitudes close to V = 30. However, all neutron stars observed
so far display optical continuum above the Rayleigh-Jeans tail of the soft
X-ray  thermal component.  This is not unexpected since black body fits tend to
overestimate \Teff . Furthermore, a non thermal optical component has been
detected in at least two cases, Geminga and PSR B0656+14. Scaling the optical
flux with the PSPC count rate of \rxa \ and \rxb \ and neglecting any
temperature effects yields V$\sim$27.2 for \rx . On the other hand, if \rx \ is
similar to Geminga or PSR B0656+14 its B magnitude could be as bright as our
limit of 26. Therefore the source may well be bright enough to be optically
identified with current means and optical imaging could allow the detection of
proper motion which is a crucial test for determining the X-ray powering
mechanism. 

\section{Conclusions}

The only optical object detected in the small HRI error circle is a late M
star most probably unrelated to the X-ray source. Altogether, X-ray and optical
observations of \rx \ strongly suggest that the soft X-ray source is due to
thermal emission from a nearby isolated neutron star. However,  based on the
presently available data,  it is not possible to distinguish between accretion
from the interstellar medium or cooling as the main X-ray emitting mechanism. 

The constancy of the X-ray flux on various time scales and the difficulties
encountered by the accretion model as a result of the small mean ambient
densities are arguments, although not fully compelling, for a cooling neutron
star. 

The undetermined neutron star spin period and the lack of sensitive
measurement of a hard X-ray component prevent us from drawing any firm
conclusion on the nature of the source. One possibility is that \rx \ is
a twin of PSR B0656+14  but that the radio beam does not intercept the earth.
Alternatively, \rx \ could be a magnetar, maybe similar to \rxa .

Further sensitive optical and X-ray observations with the XMM and AXAF
satellites could help to unveil the real nature of this object.

\begin{acknowledgements}
The ROSAT project is supported by the German Bundesministerium
f\"ur Bildung, Wissenschaft, For\-schung und Technologie (BMBF/DLR) and the
Max-Planck-Gesellschaft. 
\end{acknowledgements}

\end{document}